\documentclass[letterpaper,12pt]{article}\usepackage[]{graphicx}\usepackage[]{color}
\makeatletter
\def\maxwidth{ %
  \ifdim\Gin@nat@width>\linewidth
    \linewidth
  \else
    \Gin@nat@width
  \fi
}
\makeatother

\definecolor{fgcolor}{rgb}{0.345, 0.345, 0.345}

\usepackage{framed}
\makeatletter
 {\par\unskip\endMakeFramed%
 \at@end@of@kframe}
\makeatother

\definecolor{shadecolor}{rgb}{.97, .97, .97}
\definecolor{messagecolor}{rgb}{0, 0, 0}
\definecolor{warningcolor}{rgb}{1, 0, 1}
\definecolor{errorcolor}{rgb}{1, 0, 0}

\usepackage{alltt}

\usepackage{mathptmx}

\usepackage{graphics}

\usepackage[authoryear,comma,longnamesfirst,sectionbib]{natbib} 

\usepackage{tikz}
\usetikzlibrary{trees}
\usepackage{colortbl}
\usepackage{color}
\usepackage{booktabs}
\usepackage{hyperref}
\usepackage{amsmath}
\usepackage{amssymb}
\usepackage{float}
\usepackage{subfig}
\usepackage{multirow}
\usepackage{geometry}

\definecolor{lightgray}{rgb}{0.75, 0.75, 0.75}

\newcommand{\blanco}[1]{}

\IfFileExists{upquote.sty}{\usepackage{upquote}}{}
\begin{document}
\title{Prediction of the 2019 IHF World Men's Handball Championship -- An underdispersed sparse count data regression model}

\author{Andreas Groll
\thanks{Faculty of Statistics, TU Dortmund University, Vogelpothsweg 87, 44227 Dortmund, Germany, \emph{groll@statistik.tu-dortmund.de}}
\and Jonas Heiner
\thanks{\emph{jonas.heiner@tu-dortmund.de}}
\and Gunther Schauberger
\thanks{Chair of Epidemiology, Department of Sport and Health Sciences, Technical University of Munich, \emph{g.schauberger@tum.de}}
\and J\"orn Uhrmeister
\thanks{Faculty of Sports Sciences, Ruhr-University Bochum, 
\emph{joern.uhrmeister@rub.de}}}
\date{}

\maketitle

\noindent \textbf{Abstract} In this work, we compare several different modeling approaches for  count data applied to the scores of handball matches with regard to their predictive performances
based on all matches from the four previous IHF World Men's Handball Championships 2011 -- 2017: {\em (underdispersed) Poisson regression models},  {\em Gaussian response models} and {\em negative binomial models}. All models are based on the teams' covariate information. Within this comparison, the Gaussian response model turns out to be the best-performing prediction method on the training data and is, therefore, chosen as the final model. Based on its estimates, the IHF World Men's Handball Championship 2019 is simulated repeatedly and winning probabilities are obtained for all teams. The model clearly favors Denmark before France. Additionally, we provide survival probabilities for all teams and at all tournament stages as well as probabilities for all teams to qualify for the main round.
\bigskip


\section{Introduction}
Handball, a popular sport around the globe, is particularly important in Europe and South America.
As there are many different aspects that can be analyzed, in the last years handball had also raised an increasing interests among researchers. For example, in \cite{UhrBro:18} a group of statisticians and sports scientists selected 59 items from the play-by-play reporting of all games of the 2017 IHF World Men's Handball Championship and the involved players were compared based on their individual game actions independently of game systems, concepts and tactical tricks. The data were clustered and collected in a matrix, to add up to a ``PlayerScore". 
In another scientific work, the activity profile of elite adolescent players during
regular team handball games was examined and the physical and
motor performance of players between the first and second
halves of a match were compared \citep{CheEtAl:2011}.

In this project we elaborate on a statistical model to evaluate the chances for all teams to become champion of the upcoming IHF Handball World Cup 2019 in Denmark and Germany. For this purpose, we launched a collaboration of professional statisticians and handball experts.  While this task is rather popular for soccer (see, e.g., \citealp{GroSchTut:2015} or \citealp{Zeil:2014}), to the best of our knowledge this idea is new in handball.
In the following, we will compare several (regularized) regression approaches modeling the number of goals the two competing handball teams score in a match regarding their predicitve performances. We start with the classical model for count data, namely the Poisson regression model. Next, we allow for under- or overdispersion, where the latter can be captured by the {\em negative binomial model}. Furthermore, as for large values of the Poisson mean $\lambda$ the corresponding Poisson distribution converges to a Gaussian distribution (with $\mu=\sigma^2=\lambda$) due to the central limit theorem, this inspired us to also apply a {\em Gaussian response model}. 
Through this comparison, a best-performing model is chosen
using the mathces of the IHF World Cups 2011 -- 2017 as the training data. Based on its estimates, the IHF World Cup 2019 is
simulated repeatedly and winning probabilities are calculated for all teams. 

The rest of the manuscript is structured as follows: in Section~\ref{sec:data} 
we describe the underlying (training) data set covering (almost) all matches of 
the four preceding IHF World Cups 2011 -- 2017. Next, in Section~\ref{sec:methods} we briefly explain
four different regression approaches and compare them based on their predictive performance on the training data set. The best-performing model is then fitted to the data and used to simulate and forecast the IHF World Cup 2019 in Section~\ref{sec:prediction}.
Finally, we conclude in Section~\ref{sec:conclusion}

\section{Data}\label{sec:data}

In this section, we briefly describe the underlying data set covering all
matches of the four preceding IHF World Men's Handball Championships 2011 -- 2017 together with several
potential influence variables\footnote{Principally, a larger data set containing more IHF World Cups together with the below-mentioned covariate information could have been constructed. However, for World Cups earlier than 2011 these data were much harder or impossible to find. For this reason we rerstrict the present analysis on the four IHF World Cup 2011 -- 2017.}. Basically, we use a similar set of
covariates as \citet{GroSchTut:2015} do for their soccer FIFA World Cup analysis, with certain modifications that are necessary for handball. For each participating team, 
the covariates are observed either for the year of the respective World Cup 
(e.g.,\ GDP per capita) or shortly before the start of the World Cup (e.g.,\ a team's IHF ranking), and, therefore, vary from one World Cup to another.

Some of the variables contain information about the recent performance and sportive success of national teams, as the current form of a national team should have an influence on the team's success in the upcoming tournament. Beside these sportive variables, also certain economic factors as well as variables describing the structure of a team's squad are collected. We shall now describe in more detail these variables.

\begin{description}
\item \textbf{Economic Factors:}

\begin{description}
\item[\it GDP per capita.] To account for the general 
	increase of the gross domestic product (GDP) during 2011 -- 2017, a ratio of the GDP per capita of the respective country and the worldwide average GDP per capita is used (source: \url{http://www.imf.org/external/pubs/ft/weo/2018/01/weodata/index.aspx}).
\item[\it Population.] The population size is 
	used in relation to the respective global population to account for the general world population growth during 2011 -- 2017 (source: \url{https://population.un.org/wpp/Download/Standard/Population/}).
\end{description}\bigskip

\item \textbf{Sportive factors:}

\begin{description}
\item[\it ODDSET probability.] We convert bookmaker odds provided by the German state betting agency ODDSET into winning probabilities. The variable hence reflects the probability for each team to win the respective World Cup.
\item[\it IHF ranking.] The IHF ranking is a ranking table of national handball federations published by the IHF (source: \url{http://ihf.info/en-us/thegame/rankingtable}). The full ranking includes results of men's, women's as well as junior and youth teams and even beach handball. The points a team receives are determined from the final rankings of World Cups of the respective sub-groups and Olympic games and strictly increase over the years, so the ranking system displays an all-time ranking of the national federations. All those results can be regarded totaled or separated for each team's section. Since this project only examines men's World Cups, merely the men's ranking table will be further disposed. 
\item[\it IHF points.] In addition to the IHF ranking, we also include the precise number of IHF points the ranking is based on. This provides an even more exact all-time ranking of the national federations' historic performances.
\end{description}\bigskip

\item \textbf{Home advantage:}

\begin{description}
\item[\it Host.] It can be assumed that the host of a Word Cup might have a home advantage, since the players' experience a stronger crowd support in the arena and are more conversant with the host country's cultural circumstances. Hence, a dummy is included indicating if a national team is a hosting country. Since the World Cup 2019 is jointly hosted by Germany and Denmark, both are treated equally.
\item[\it Continental federation.] The IHF is the parent organization of the different continental federations, the African Handball Confederation (CAHB), the Asian Handball Federation (AHF), the European Handball Federation (EHF), the 
Oceania Continent Handball Federation (OCHF) and the Pan-American Team Handball Federation (PATHF).

The nation's affiliation to the same continental federation as the host could on the one hand influence the team's performance similar to the Word Cup's host by their better habituation with the host's conventions. Additionally, supporters of those teams have a shorter arrival. On the other, hand handball is not equally prevalent on every continent, especially European club handball is most popular. To capture potential performance differences between the continental federations, two variables are added to the data set. A dummy determining whether a nation is located in {\it Europe}, and a dummy seizing whether a nation belongs to the {\it same umbrella organization as the Word Cup host}.
\end{description}
\bigskip

\item \textbf{Factors describing the team's structure:}

	The following variables describe the structure of the teams. 
	They were observed with the 16-player-squad 
	nominated for the respective World~Cup.\medskip 

\begin{description}
\item[\it (Second) maximum number of teammates.] For each squad, both the maximum 
	and second maximum number of teammates playing together in the same national club are counted.
\item[\it Average age.] The average age of each squad is collected. 
However, very young players might be rather inexperienced at big tournaments and some older players might lack a bit concerning their condition. For this reason we assume an ideal athlete's age, here represented by the average age of all squads that participated in World Cups throughout the last eight years, so that the absolute divergence between a national team's average age and that ideal age is surveyed. 
\item[\it Average height.] The average height of a team can possibly impact the team's power. Tall players might have an advantage over short players, as they can release a shot on goal above a defender more easily. Therefore, we include the team's average height in meters.
\item[\it Number of EHF Champions League (EHF-cup) players.] 	
As club handball is mainly based on the European continent, the EHF Champions League is viewed as the most attractive competition, as numerous of the best club teams in the world participate and only the best manage to reach the final stages of the competition. Hence, also the best players play for these clubs. For this reason we include the number of players of each country that reached the EHF Champions League semifinals in the previous year of the respective World Cup. The same data is collected for the second biggest European club competition, the EHF-cup.
\item[\it Number of players abroad/Legionnaires.] For each squad, the number of players 
	playing in clubs abroad is counted.
\end{description}\bigskip

\item \textbf{Factors describing the team's coach:} 

The players of course extinguish the most important part of a squad, but every team additionally needs an eligible coach to instruct the players. Therefore, some observable trainer characteristics are gathered, namely \textit{Age} and \textit{Tenure} of the coach plus a dummy variable that determines whether he shares the same \textit{Nationality} as his team.
\end{description}

\noindent In total, this adds up to 18 variables which were collected separately for each World Cup and each participating team. As an illustration, Table~\ref{data1} shows the results (\ref{tab:results}) and (parts of) the covariates (\ref{tab:covar}) of the respective teams, exemplarily for the first four matches of the IHF World Cup 2011. We use this data excerpt to illustrate how the final data set is constructed.

	\begin{table}[h]
\small
\caption{\label{data1} Exemplary table showing the results of four matches and parts of the covariates of the involved teams.}
\centering
\subfloat[Table of results \label{tab:results}]{
\begin{tabular}{lcr}
  \hline
 &  &  \\ 
  \hline
FRA \includegraphics[width=0.4cm]{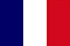} & 32:19 &  \includegraphics[width=0.4cm]{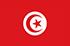} \;TUN\\
ESP \includegraphics[width=0.4cm]{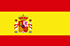} & 33:22 &  \includegraphics[width=0.4cm]{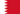} \;BAH\\
BAH \includegraphics[width=0.4cm]{BAH.png} & 18:38 &  \includegraphics[width=0.4cm]{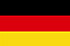} \;GER\\
TUN \includegraphics[width=0.4cm]{TUN.png} & 18:21 &  \includegraphics[width=0.4cm]{ESP.png} \;ESP\\
  \vdots & \vdots & \vdots  \\
  \hline
\end{tabular}}
\hspace*{0.8cm}
\subfloat[Table of (original) covariates \label{tab:covar}]{
\begin{tabular}{llrrrrr}
  \hline
World Cup & Team &  Age & Rank & Oddset &   \ldots \\ 
  \hline
2011 & France  & 29.0 & 5 & 0.291 & \ldots \\ 
2011 &  Tunisia & 26.4 & 17 & 0.007 & \ldots \\ 
2011 &  Germany & 26.9 & 1 & 0.007 & \ldots\\ 
2011 &  Bahrain & 29.0 & 48 & 0.001 & \ldots\\ 
2011 &  Spain & 26.8 & 7 & 0.131 & \ldots\\ 
  \vdots & \vdots & \vdots & \vdots & \vdots  &  $\ddots$ \\
   \hline
\end{tabular}
}
\end{table}

\noindent For the modeling techniques that we shall introduce in the following sections, all of the metric covariates are incorporated in the form of differences. For example, the final variable {\it Rank} will be the difference between the IHF ranks of both teams. The categorical variables {\it Host},
{\it Nationality} as well as the two continental federation variables, however, are included as separate variables for both competing teams.
For the variable {\it Host}, for example, this results in two columns of the corresponding design matrix denoted by 
{\it Host} and {\it Host.Oppo}, where {\it Host} is indicating whether the first-named team
is a World Cup host and {\it Host.Oppo} whether its opponent is.

As we use the number of goals of each team directly as the response variable, each match corresponds to two different observations, one per team. For the covariates, we consider differences which are computed from the perspective of the first-named team. For illustration, the resulting final data structure for the exemplary matches from Table~\ref{data1} is displayed in Table~\ref{data2}.

\begin{table}[!h]
\small
\centering
\caption{Exemplary table illustrating the data structure.}\label{data2}
\begin{tabular}{rllrrrr}
  \hline
Goals & Team & Opponent & Age & Rank & Oddset &  ... \\ 
  \hline
32 & France & Tunisia & 0.81 & 12 & 0.284 &  ...  \\ 
  19 & Tunesia & France & - 0.81 & -12 & -0.284 &  ...  \\ 
  33 & Spain & Bahrain & 1.21 & -41 & 0.129 &  ...  \\ 
  22 & Bahrain & Spain & -1.21 & 41 & -0.129 &  ...  \\ 
    18 & Bahrain & Germany & 0.10 & 47 & -0.064 &  ... \\ 
  38 & Germany & Bahrain & -0.10 & -47 & 0.064 &  ... \\ 
  18 & Tunisia & Spain & -0.81 & 10 & -0.124 &  ...  \\ 
  21 & Spain & Tunisia & 0.81 & -10 & 0.124 &  ... \\ 
	 \vdots & \vdots & \vdots & \vdots & \vdots & \vdots &  $\ddots$ \\
   \hline
\end{tabular}
\end{table}

\noindent Due to some missing covariate values for a few games, altogether the final data set contains 334 out of 354 matches from the four handball World Cups 2011 -- 2017. Note that in all the models described in the next section, we incorporate all of the above mentioned covariates. However, not all of them will be selected by the introduced penalization technique. Instead, rather sparse models will be prefered.

\section{Methods}\label{sec:methods}

In this section, we briefly describe several different regression approaches 
that generally come into consideration when the goals scored in single handball matches are 
directly modeled. Actually, most of them (or slight modifications thereof) 
have already been used in former research on soccer data and, generally, all yielded satisfactory results. However, some adjustments are necessary for handball. 
All methods described in this section can be directly applied to data in the format of Table~\ref{data2} from Section~\ref{sec:data}. Hence, each score is treated as a single observation and one obtains two observations per match.
We aim to choose the approach that has the best performance regarding prediction and then use it to predict the IHF World Men's Handball Championship 2019.

\subsection{Poisson model}\label{subsec:pois}

A traditional approach which is often applied, for example, to model soccer results is based on Poisson regression. In this case, the scores of the competing teams are treated as (conditionally) independent variables following a Poisson distribution (conditioned on certain covariates), as introduced in the seminal works of \citealp{Mah:82} and \citealp{DixCol:97}.  

As already stated, each score from a match of two handball teams is treated as a single observation. Accordingly, similar to the regression approach investogated in \cite{GroEtAl:WM2018}, for $n$ teams the respective model has the form
\begin{eqnarray}
Y_{ijk}|\boldsymbol{x}_{ik},\boldsymbol{x}_{jk}&\sim &Po(\lambda_{ijk})\,,\nonumber\\ 
\label{lasso:model}\log(\lambda_{ijk})&=&\eta_{ijk}\,:=\,\beta_0 + (\boldsymbol{x}_{ik}-\boldsymbol{x}_{jk})^\top\boldsymbol{\beta}+\boldsymbol{z}_{ik}^\top\boldsymbol{\gamma}+\boldsymbol{z}_{jk}^\top\boldsymbol{\delta}\,,
\end{eqnarray}
where $Y_{ijk}$ denotes the score of handball team $i$ against team $j$ in tournament $k$ with $i,j\in\{1,\ldots,n\},~i\neq j$ and $\eta_{ijk}$ is the corresponding linear predictor. The metric characteristics of both competing teams are captured in the $p$-dimensional vectors $\boldsymbol{x}_{ik}, \boldsymbol{x}_{jk}$, while $\boldsymbol{z}_{ik}$ and $\boldsymbol{z}_{jk}$ capture dummy variables for the categorical covariates {\it Host},
{\it Nationality} as well as the two continental federation variables (built, for example, by reference encoding), separately for the considered teams and their respective opponents. For these variables, it is not sensible to build differences between the respective values. Furthermore, $\boldsymbol{\beta}$ is a parameter vector which captures the linear effects of all metric covariate differences and $\boldsymbol{\gamma}$ and $\boldsymbol{\delta}$ collect the effects of the dummy variables corresponding to the teams and their opponents, respectively. For notational convenience, we collect all covariate effects in the $\tilde p$-dimensional real-valued vector $\boldsymbol{\theta}^\top=(\boldsymbol{\beta}^\top, \boldsymbol{\gamma}^\top, \boldsymbol{\delta}^\top)$. 

If, as in our case, several covariates of the competing teams are included into the model it is sensible to use regularization techniques when estimating the models to allow for variable selection and to avoid overfitting. In the following, we will introduce such a basic regularization approach, namely the conventional Lasso (least absolute shrinkage and selection operator; \citealp{Tibshirani:96}).
For estimation, instead of the regular likelihood $l(\beta_0,\boldsymbol{\theta})$ the penalized likelihood 
\begin{eqnarray}\label{eq:lasso}
l_p(\beta_0,\boldsymbol{\theta}) = l(\beta_0,\boldsymbol{\theta}) - \xi P(\beta_0,\boldsymbol{\theta})
\end{eqnarray}
is maximized, where $P(\beta_0,\boldsymbol{\theta})=\sum_{v=1}^{\tilde p}|\theta_v|$ denotes the ordinary Lasso penalty with tuning parameter $\xi$. The optimal value for the tuning parameter $\xi$ will be determined by 10-fold cross-validation (CV). The model will be fitted using the function \texttt{cv.glmnet} from the \texttt{R}-package \texttt{glmnet} \citep{FrieEtAl:2010}. In contrast to the similar ridge penalty \citep{HoeKen:70}, which penalizes squared parameters instead of absolute values, Lasso does not only shrink parameters towards zero, but is able to set them to exactly zero. Therefore, depending on the chosen value of the tuning parameter, Lasso also enforces variable selection. 

\subsection{Overdispersed Poisson model / negative binomial model}\label{subsec:NegBin}

The Poisson model introduced in the previous section is built on the rather strong assumption
$E\left[Y_{ijk}|\boldsymbol{x}_{ik},\boldsymbol{x}_{jk}\right] = Var\left(Y_{ijk}|\boldsymbol{x}_{ik},\boldsymbol{x}_{jk}\right) = \lambda_{ijk}$, i.e.\ that the expectation of the distribution equates the variance. For the case of World Cup handball matches, the  (marginal) average number of goals is around  30 (for example, $\bar{y} = 27.33$ for the matches of the IHF World Cups 2011 -- 2017) and supposably the corresponding variance could differ substantially. 

A case often treated in the literature is the case when $Var(Y)>E[Y]$, the so-called overdispersion. But for handball matches, also the contrary could be possible, namely that $Var(Y)<E[Y]$ holds. In both cases, one typically assumes that $Var(Y)=\phi\cdot E[Y]$ holds, where $\phi$ is called {\it dispersion parameter} and can be estimated via

\begin{equation}\label{eq:dispersion}
\hat\phi = \frac{1}{N-df} \sum\limits_{i=1}^N r_i^2,
\end{equation}
where $N$ is the number of observations and $r_i$ the model's Pearson residuals.

We will first focus on the (more familiar) case of overdispersion. It is well known that the overdispersed Poisson model can be obtained by using the negative binomial model. To combine this model class with the Lasso penalty from equation~\eqref{eq:lasso}, the \texttt{cv.glmregNB} function from the R-package \texttt{mpath} \citep{Zhu:2018} can be used (see also, for example, \citealp{ZhuEtAl:2018}).

\subsection{Underdispersed Poisson model}\label{subsec:under_pois}

If we fit the (regularized) Poisson model from Section~\ref{subsec:pois} to our IHF World Cup data and then estimate the dispersion parameter via equation~\eqref{eq:dispersion}, we obtain a value for $\hat\phi$ clearly smaller than one ($\hat\phi=0.74$), i.e.\ substantial underdispersion. Hence, the variance of the goals in IHF World Cup matches seems to be smaller than their mean.

To be able to simulate from an underdispersed Poisson distribution (which we would need later on to simulate matches from the IHF World Cup 2019), the \texttt{rdoublepois} function from the \texttt{rmutil}-package (\citep{SwiLin:2018}) can be used.

\subsection{The Gaussian response model}\label{subsec:gauss}
It is well-known that for large values of the Poisson mean $\lambda$ the corresponding Poisson distribution converges to a Gaussian distribution (with $\mu=\sigma^2=\lambda$) due to the central limit theorem. In practice, for values $\lambda\approx 30$ (or larger) the approximation of the Poisson via the Gaussian distribution is already quite satisfactory.
As we have already seen in Section~\ref{subsec:NegBin} that the average number of goals in handball World Cup matches is close to 30, this inspired us to also apply a Gaussian response model.

However, instead of forcing the mean to equate the variance, we again allow for $\mu\neq\sigma^2$, i.e.\ for potential (constant) over- or underdispersion. Note here that the main difference to the over- and underdispersion models from the two preceding sections is that there each observation obtains its own variance via $Var\left(Y_{ijk}|\boldsymbol{x}_{ik},\boldsymbol{x}_{jk}\right) = \hat\phi\cdot\lambda_{ijk}$, where in the Gaussian response model all observations have the same 
variance $\hat\sigma^2$. On our World Cup 2011 -- 2017 data, we obtain $\hat\sigma^2=20.13$, which compared to the average number of goals $\bar{y} = 27.33$ indicates a certain amount of (constant) underdispersion.

We also want to point out here that in order to be able to simulate a precise match result from the
model's distribution (and then, successively, to calculate probabilities for the three match results {\it win}, {\it draw} or {\it loss}), we round results to the next natural number. In general, the Lasso-regularized Gaussian response model will again be fitted using the function \texttt{cv.glmnet} from the \texttt{R}-package \texttt{glmnet} based on the linear predictor $\eta_{ijk}$ defined in equation~\eqref{lasso:model}.

\subsection{Increase model sparsity}\label{subsec:sparse}

Note that in addition to the conventional Lasso solution minimizing the 10-fold CV error, a second, sparser solution could be used. Here, the optimal value for the tuning parameter $\xi$ is chosen by a different strategy: instead of choosing the model with the minimal CV error the most restrictive model is chosen which is within one standard error of the minimum of the CV error. 
While it is directly provided by the \texttt{cv.glmnet} function from the \texttt{glmnet} package, for the \texttt{cv.glmregNB} function it had to be calculated manually.
In the following section, where all the different models from above are compared, for each model class also this sparser solution is calculated and included in the comparison.

\subsection{Comparing methods}\label{subsec:compare}

The four different approaches introduced in Sections~\ref{subsec:pois} - \ref{subsec:gauss} are now compared with regard to their predictive performance. For this purpose, we apply the following general procedure to the World Cup 2011 -- 2017 data which had already been applied to soccer World Cup data in \cite{GroEtAl:WM2018}: 
\begin{enumerate}{\it
\item Form a training data set containing three out of four World Cups.\vspace{0.1cm}
\item Fit each of the methods to the training data.\vspace{0.1cm}
\item Predict the left-out World Cup using each of the prediction methods.\vspace{0.1cm}
\item Iterate steps 1-3 such that each World Cup is once the left-out one.\vspace{0.1cm}
\item Compare predicted and real outcomes for all prediction methods.}\vspace{-0.1cm}
\end{enumerate}
This procedure ensures that each match from the total data set is once part of the test data and we obtain out-of-sample predictions for all matches. In step~{\it 5}, several different performance measures for the quality of the predictions are investigated.

Let $\tilde y_i\in\{1,2,3\}$ be the true ordinal match outcomes for all $i=1,\ldots,N$ matches  from the four considered World Cups. Additionally, let $\hat\pi_{1i},\hat\pi_{2i},\hat\pi_{3i},~i=1,\ldots,N$, be the predicted probabilities for the match outcomes obtained by one of the different methods introduced in Sections~\ref{subsec:pois} - \ref{subsec:gauss}. Further, let $G_{1i}$ and $G_{2i}$ denote the random variables representing the number of goals scored by two competing teams in match $i$. Then, the probabilities $\hat \pi_{1i}=P(G_{1i}>G_{2i}), \hat \pi_{2i}=P(G_{1i}=G_{2i})$ and $\hat \pi_{3i}=P(G_{1i}<G_{2i})$ can be computed/simulated
based on the respective underlying (conditionally) independent response distributions $F_{1i},F_{2i}$ with $G_{1i}\sim F_{1i}$ and $G_{2i}\sim F_{2i}$. The two distributions $F_{1i},F_{2i}$ depend on the corresponding linear predictors $\eta_{ijk}$ and $\eta_{jik}$ from equation~\eqref{lasso:model}.

Based on these predicted probabilities, following \cite{GroEtAl:WM2018} we use three different performance measures to compare the predictive power of the methods:
\begin{itemize}
\item the multinomial {\it likelihood}, which for a single match outcome is defined as $\hat \pi_{1i}^{\delta_{1\tilde y_i}} \hat \pi_{2i}^{\delta_{2\tilde y_i}} \hat \pi_{3i}^{\delta_{3 \tilde y_i}}$, with $\delta_{r\tilde y_i}$ denoting Kronecker's delta. It  reflects the probability of a correct prediction. Hence, a large value reflects a good fit.\vspace{0.1cm}
\item  the {\it classification rate}, based on the indicator functions $\mathbb{I}(\tilde y_i=\underset{r\in\{1,2,3\}}{\mbox{arg\,max }}(\hat\pi_{ri}))$, indicating whether match $i$ was correctly classified.
Again, a large value of the classification rate reflects a good fit.\vspace{0.1cm}
\item the {\it rank probability score} (RPS) which, in contrast to both measures introduced above, explicitly accounts for the ordinal structure of the responses. 
For our purpose, it can be defined as $\frac{1}{3-1} \sum\limits_{r=1}^{3-1}\left( \sum\limits_{l=1}^{r}(\hat\pi_{li} - \delta_{l\tilde y_i})\right)^{2}$. As the RPS is an error measure, here a low value represents a good fit.
\end{itemize}
Odds provided by bookmakers serve as a natural benchmark for these predictive performance measures. For this purpose, we collected the so-called ``three-way'' odds for (almost) all matches of the IHF World Cups 2011 -- 2017\footnote{Three-way odds consider only the match tendency with possible results \emph{victory team 1}, \emph{draw} or \emph{defeat team 1} and are usually fixed some days before the corresponding match takes place. The three-way odds were obtained from the website \url{https://www.betexplorer.com/handball/world/}.}.
By taking the three quantities $\tilde \pi_{ri}=1/\mbox{odds}_{ri}, r\in\{1,2,3\}$, of a match $i$ and by normalizing with $c_i:=\sum_{r=1}^{3}\tilde \pi_{ri}$ in order to adjust for the bookmaker's margins, the odds can be directly transformed into probabilities using $\hat \pi_{ri}=\tilde \pi_{ri}/c_i$
\footnote{The transformed probabilities implicitely assume that the bookmaker's margins are equally distributed on the three possible match tendencies.}.

As we later want to predict both winning probabilities for all teams and the whole tournament course for the IHF World Cup 2019,
we are also interested in the performance of the regarded methods with respect to the prediction of the exact number of goals. In order to identify the teams that qualify during both group stages, the precise final group standings need to be determined. To be able to do so, the precise results of the matches in the group 
stage play a crucial role\footnote{\label{foot:mode}The final group standings are determined by (1) the number of points, (2) head-to-head points (3) head-to-head goal difference, (4) head-to-head number of goals scored, (5) goal difference and (6) total number of goals. If no distinct decision can be taken, the decision is taken by lot.}.

For this reason, we also evaluate the different regression models' performances
with regard to the quadratic error between the observed and predicted
number of goals for each match and each team, as well as between the observed and predicted goal difference. Let now $y_{ijk}$, for $i,j=1,\ldots,n$ and $k\in\{2011,2013,2015,2017\}$,
denote the observed numbers of goals scored by team $i$ against team $j$ in tournament $k$ and
$\hat y_{ijk}$ a corresponding predicted value, obtained by one of the models from Sections~\ref{subsec:pois} - \ref{subsec:gauss}. 
Then we calculate the two quadratic errors $(y_{ijk}-\hat y_{ijk})^2$ and $\left((y_{ijk}-y_{jik})-(\hat y_{ijk}-\hat y_{jik})\right)^2$ for all $N$ matches of the four IHF World Cups 2011 -- 2017. Finally, per method we calculate (mean) quadratic errors. 

Table~\ref{tab:probs_old} displays the results for these five performance measures 
for the models introduced in Sections~\ref{subsec:pois} - \ref{subsec:gauss} as well as for the bookmakers, averaged over 334 matches from the four IHF World Cups 2011 -- 2017. While the bookmakers serve as a benchmark and yield the best results with respect to all ordinal critera, the second-best method's results are highlighted in bold text. It turns out that the Poisson and the underdispersed Poisson model yield very good results with respect to the classification rate, while the Gaussian response model is (in some cases clearly) the best performer regarding all other criteria. As no overdispersion (and, actually, underdispersion) is found in the data, the negative binomial model's results are almost indistinguishable from those of the (conventional) Poisson model. 
The more sparse Lasso estimators introduced in Section~\ref{subsec:sparse} perform substantially worse in terms of prediction accuracy compared to the conventional Lasso solution.

Based on these results, we chose the regularized Gaussian response model with constant (and rather low) variance as our final model and shall use it in the next section to simulate the IHF World Cup 2019.

\begin{table}[H]
\small
\caption{\label{tab:probs_old}Comparison of the different methods for ordinal match outcomes; the second-best method's results are highlighted in bold text.}\vspace{0.2cm}

\centering
\begin{tabular}{lrrrrr}
  & Multinomial & Class. Rate & RPS & Goals & Goal Difference \\ 
  \toprule
Pois & 0.6271 & 0.7665 & 0.1546 & 22.4944 & 39.1713 \\ 
   \midrule
Pois ($\lambda_{1se}$) & 0.5952 & 0.7365 & 0.1627 & 22.5759 & 39.8042 \\ 
   \midrule
underdis. Pois & 0.6409 & {\bf 0.7665} & 0.1526 & 22.4944 & 39.1713 \\ 
   \midrule
underdis. Pois ($\lambda_{1se}$) & 0.6047 & 0.7335 & 0.1598 & 22.5759 & 39.8042 \\ 
   \midrule
NB & 0.6285 & 0.7575 & 0.1546 & 22.4836 & 39.2347 \\ 
   \midrule
NB ($\lambda_{1se}$) & 0.6024 & 0.7455 & 0.1592 & 22.3320 & 38.6094 \\ 
   \midrule
Gauss & {\bf 0.6413} & 0.7575 & {\bf 0.1512} & {\bf 22.0603} & {\bf 38.0023} \\ 
   \midrule
Gauss ($\lambda_{1se}$) & 0.6055 & 0.7365 & 0.1598 & 22.5894 & 39.7949 \\ 
   \midrule
Odds & 0.6688 & 0.8114 & 0.1256 & - & - \\ 
   \bottomrule
\end{tabular}

\end{table}


\section{Prediction of the IHF World Cup 2019}\label{sec:prediction}

Now we apply the best-performing model from Section~\ref{sec:methods}, namely the regularized Gaussian response model with constant underdispersion, to the full World Cup 2011 -- 2017 training data and will then use it to calculate winning probabilities for the World Cup 2019. For this purpose, the covariate information from Section~\ref{sec:data} has to be collected for all teams participating at the 2019 World~Cup. 

It has to be stated that at the time this analysis has been performed, namely at the first tournament day (June 10, 2019) right before the start of the tournament, the teams of Bahrein and Sweden had listed squads consisting of 15 players only and there will probably be one more player moving up soon. Hence, for those two teams the covariates corresponding to natural-numbered team structure variables (such as, e.g., the {\it number of legionnaires}) have been normalized to be comparable to 16-player squads by multiplying them with the factor $16/15$. Another special case concerns the team of Korea. As this team is fromed by a selection of players from both South and North Korea, the federation was given the special approval to nominate 20 players. As it actually might be an advantage to have a larger squad we abstained here from normalizing the covariate values from the Korean team.

\begin{figure}
\includegraphics[width=0.55\textwidth]{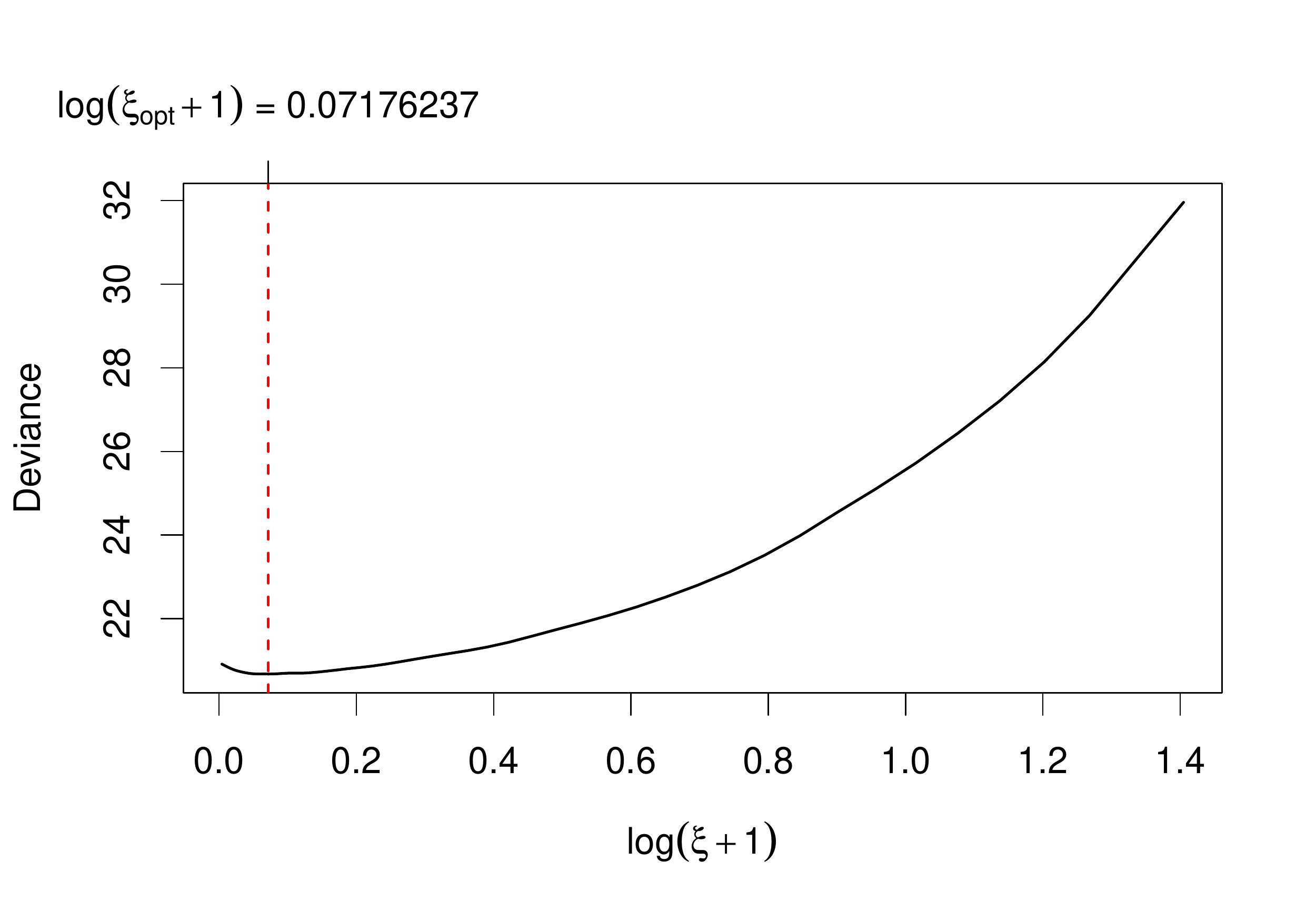}\hspace{-0.5cm}
\includegraphics[width=0.55\textwidth]{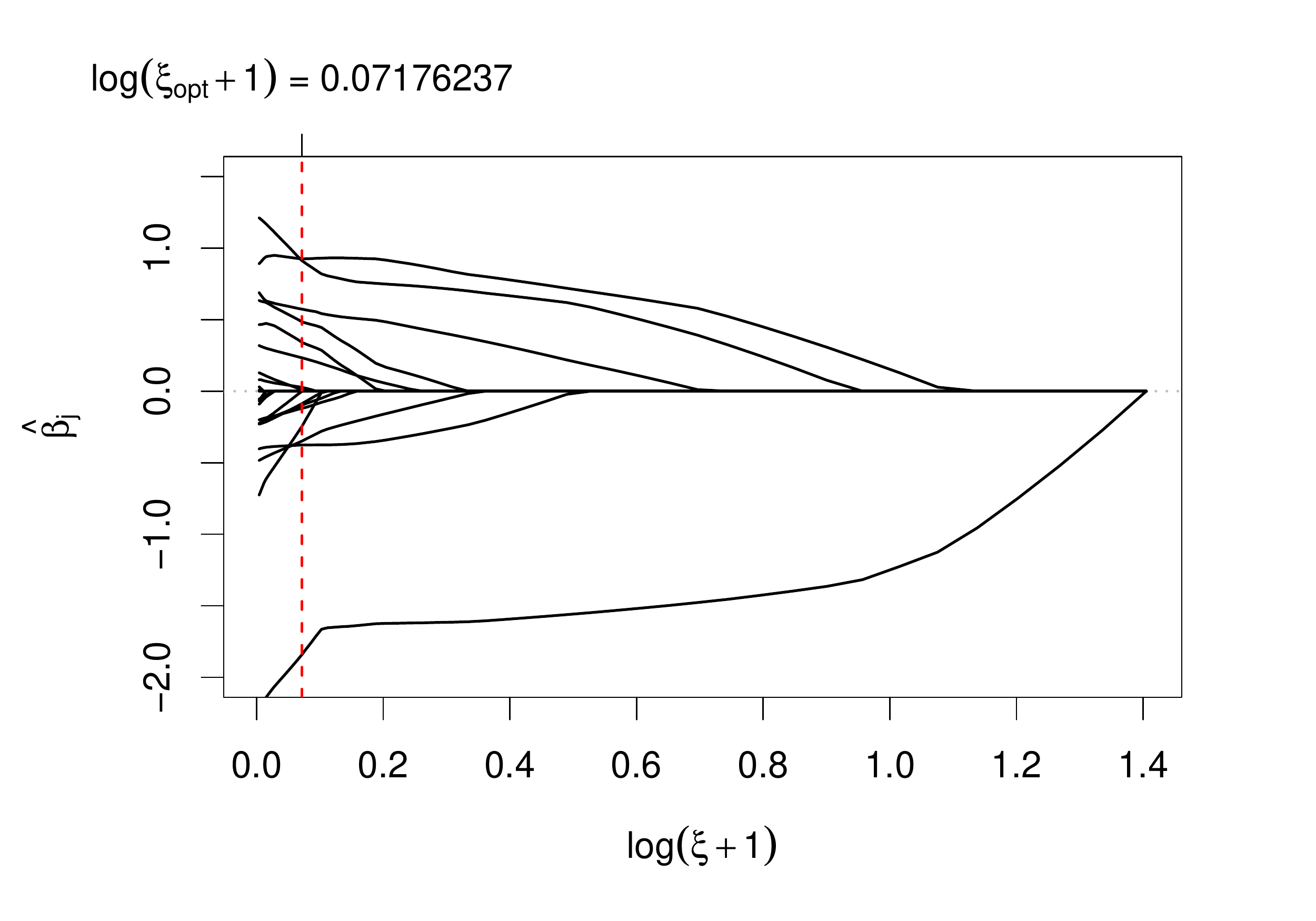}\vspace{-0.3cm}
\caption{Left: 10-fold CV deviance for Gaussian response model on 
IHF World Cup data 2011 -– 2017; Right: Coefficient paths vs. (logarithmized) penalty parameter $\xi$; optimal value of the penalty
parameter $\xi$ shown by vertical line.}\label{fig:lasso}
\end{figure}

The optimal tuning parameter $\xi$ of the L1-penalized Gaussian response model, which minimizes the deviance shown in Figure~\ref{fig:lasso} (left), leads to a model with
16 (out of possibly 22) regression coefficients different
from zero. The paths illustrated
in Figure~\ref{fig:lasso} (right) show that three covariates enter the model rather early. These are the {\it Rank}, the {\it Height} and the {\it Odds}, which seem to be rather important when determining the score in a handball World Cup match. The corresponding fixed effects estimates for the (scaled) covariates are shown in Table~\ref{tab:lasso_coefs}.

\begin{table}[H]
\small
\caption{\label{tab:lasso_coefs}Estimates of the covariate effects for the IHF World Cups 2011 -- 2017.}\vspace{0.2cm}
\centering
\begin{tabular}{rr}
 variable & estimate \\ 
  \toprule
Age & -0.3486 \\ 
  Height & 0.9243 \\ 
  Trainer.age & 0.2331 \\ 
  Trainer.tenure & -0.1202 \\ 
  Legionairs & 0.3408 \\ 
  CL.final4 & -0.0006 \\ 
  EHF.final4 & 0.0000 \\ 
  max.teammates & 0.4842 \\ 
  sec.max.teammates & 0.0000 \\ 
  Trainer.nat & -0.0973 \\ 
  Odds & 0.9117 \\ 
  ihf.points & -0.2449 \\ 
  Rank & -1.8404 \\ 
  GDP & 0.0000 \\ 
  Population & 0.0000 \\ 
  Host & -0.0868 \\ 
  Confed & 0.5734 \\ 
  Continent & 0.0266 \\ 
  Host.oppo & -0.3763 \\ 
  Trainer.nat.oppo & 0.0076 \\ 
  Confed.oppo & 0.0000 \\ 
  Continent.oppo & 0.0000 \\ 
   \bottomrule
\end{tabular}

\end{table}

Based on the estimates from Table~\ref{tab:lasso_coefs} and the covariates of all teams for the IHF World Cup 2019, 
we can now simulate all matches from the preliminary round. Next, we can simulate all resulting matches in the main round and determine those teams that reach the semi-finals and, finally, those two teams that reach the final and the World Champion. We repeat the simulation of the whole tournament 100,000 times. This way, for each of the 24 participating teams probabilities to reach the different tournament stages and, finally, to win the tournament are obtained.



\subsection{Probabilities for IHF World Cup 2019 Winner}

For each match in the World Cup 2019, the model can be used to predict an expected number of goals for both teams. Given the expected number of goals, a real result is drawn by assuming two (conditionally) independent Gaussian distributions for both scores, which are then rounded to the closest natural number. Based on these results, all 60 matches  from the preliminary round can be simulated and final group standings can be calculated. Due to the fact that real results are simulated, we can precisely follow the official IHF rules when
determining the final group standings (see footnote~\ref{foot:mode}). This enables us to determine the matches in the main round and we can continue by simulating those matches. Again, if the final group standings
are calculated, the semi-final is determined. Next, the semi-final can be simulated and the final is determined. In the case of draws in ``knockout" matches, we simulate extra-time by a second simulated result. However, here we multiply the expected number of goals by the factor 1/6 to account for the shorter time to score (10 min instead of 60 min). In the case of a further draw in the first extra-time, we repeat this procedure. If the second extra time still ends in a draw we simulate the penalty shootout by a (virtual) coin flip.

\begin{table}[!h]
\small
\caption{\label{winner_probs}Estimated probabilities (in \%) for reaching (at least) the main round or the given final ranks in the IHF World Cup 2019 for all 24 teams based on 100,000 simulation runs of the IHF World Cup together with winning probabilities based on the ODDSET odds.}\vspace{0.4cm}

\centering
\begin{tabular}{lllrrrrrrrrrr}
   \toprule
 &  &  & Main & 8th & 7th & 6th & 5th & 4th & 3rd & 2nd & Champion & Oddset \\ 
   \midrule
1. & \includegraphics[width=0.4cm]{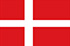} & DEN & 99.4 & 98.0 & 96.8 & 93.9 & 90.6 & 83.5 & 75.7 & 54.9 & 37.2 & 25.4 \\ 
  2. & \includegraphics[width=0.4cm]{FRA.png} & FRA & 89.1 & 80.1 & 77.9 & 70.3 & 65.8 & 54.7 & 47.3 & 33.5 & 19.4 & 23.7 \\ 
  3. & \includegraphics[width=0.4cm]{GER.png} & GER & 80.2 & 65.6 & 61.9 & 52.8 & 46.9 & 36.7 & 29.1 & 19.3 & 9.2 & 11.8 \\ 
  4. & \includegraphics[width=0.4cm]{ESP.png} & ESP & 96.0 & 75.1 & 69.9 & 58.4 & 50.9 & 39.2 & 30.2 & 19.9 & 9.0 & 14.2 \\ 
  5. & \includegraphics[width=0.4cm]{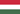} & HUN & 95.5 & 83.6 & 74.6 & 65.6 & 52.5 & 41.1 & 27.5 & 17.5 & 6.4 & 1.8 \\ 
  6. & \includegraphics[width=0.4cm]{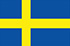} & SWE & 93.8 & 78.8 & 67.8 & 58.3 & 44.3 & 33.7 & 21.3 & 13.1 & 4.4 & 5.1 \\ 
  7. & \includegraphics[width=0.4cm]{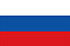} & RUS & 69.0 & 50.3 & 45.7 & 37.1 & 31.2 & 23.5 & 17.0 & 10.6 & 4.3 & 0.7 \\ 
  8. & \includegraphics[width=0.4cm]{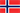} & NOR & 93.5 & 78.6 & 67.2 & 57.9 & 43.8 & 33.4 & 20.7 & 12.9 & 4.1 & 7.9 \\ 
  9. & \includegraphics[width=0.4cm]{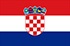} & CRO & 91.8 & 58.5 & 51.3 & 39.6 & 32.2 & 23.7 & 15.9 & 9.7 & 3.5 & 5.1 \\ 
  10. & \includegraphics[width=0.4cm]{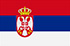} & SRB & 52.8 & 32.9 & 28.2 & 21.9 & 17.3 & 12.7 & 8.1 & 4.8 & 1.6 & 0.5 \\ 
  11. & \includegraphics[width=0.4cm]{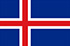} & ICE & 80.2 & 32.1 & 25.0 & 17.9 & 12.6 & 8.9 & 4.7 & 2.6 & 0.7 & 0.4 \\ 
  12. & \includegraphics[width=0.4cm]{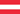} & AUT & 53.7 & 17.8 & 9.4 & 7.4 & 3.5 & 2.6 & 0.8 & 0.4 & 0.0 & 0.1 \\ 
  13. & \includegraphics[width=0.4cm]{TUN.png} & TUN & 50.7 & 15.8 & 8.2 & 6.4 & 2.9 & 2.2 & 0.5 & 0.3 & 0.0 & 0.2 \\ 
  14. & \includegraphics[width=0.4cm]{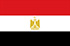} & EGY & 46.5 & 14.5 & 7.6 & 6.0 & 2.7 & 2.1 & 0.6 & 0.3 & 0.0 & 0.2 \\ 
  15. & \includegraphics[width=0.4cm]{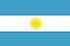} & ARG & 28.4 & 6.5 & 2.9 & 2.4 & 0.9 & 0.7 & 0.2 & 0.1 & 0.0 & 0.1 \\ 
  16. & \includegraphics[width=0.4cm]{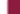} & KAT & 24.7 & 5.0 & 2.2 & 1.7 & 0.7 & 0.5 & 0.1 & 0.0 & 0.0 & 1.1 \\ 
  17. & \includegraphics[width=0.4cm]{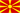} & MAC & 21.4 & 2.5 & 1.3 & 0.8 & 0.4 & 0.3 & 0.1 & 0.0 & 0.0 & 0.7 \\ 
  18. & \includegraphics[width=0.4cm]{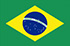} & BRA & 7.9 & 2.0 & 1.3 & 0.9 & 0.5 & 0.4 & 0.1 & 0.0 & 0.0 & 0.6 \\ 
  19. & \includegraphics[width=0.4cm]{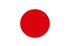} & JPN & 10.5 & 0.7 & 0.3 & 0.2 & 0.1 & 0.1 & 0.0 & 0.0 & 0.0 & 0.1 \\ 
  20. & \includegraphics[width=0.4cm]{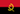} & ANG & 11.0 & 1.4 & 0.4 & 0.4 & 0.1 & 0.1 & 0.0 & 0.0 & 0.0 & 0.1 \\ 
  21. & \includegraphics[width=0.4cm]{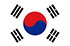} & KOR & 1.1 & 0.1 & 0.0 & 0.0 & 0.0 & 0.0 & 0.0 & 0.0 & 0.0 & 0.1 \\ 
  22. & \includegraphics[width=0.4cm]{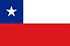} & CHI & 1.3 & 0.0 & 0.0 & 0.0 & 0.0 & 0.0 & 0.0 & 0.0 & 0.0 & 0.1 \\ 
  23. & \includegraphics[width=0.4cm]{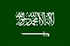} & KSA & 1.4 & 0.0 & 0.0 & 0.0 & 0.0 & 0.0 & 0.0 & 0.0 & 0.0 & 0.1 \\ 
  24. & \includegraphics[width=0.4cm]{BAH.png} & BAH & 0.2 & 0.0 & 0.0 & 0.0 & 0.0 & 0.0 & 0.0 & 0.0 & 0.0 & 0.1 \\ 
   \bottomrule
\end{tabular}

\end{table}

Following this strategy, a whole tournament run can be simulated, which we repeat 100,000
times. Based on these simulations, for each of the 24 participating
teams probabilities to reach (at least) the main round or the given final rank and,
finally, to win the tournament are obtained. These are summarized
in Table~\ref{winner_probs} together with the winning probabilities
based on the ODDSET odds for comparison.

Apparently, the resulting winning probabilities show some  
discrepancies from the probabilities based on the bookmaker's odds.
Though the upper and lower half of the teams according to our calculated probabilities 
seem to coincide quite well with the overall ranking according to the bookmaker's odds,
for single teams from the upper half, in particular, Denmark, Spain and Hungary, 
the differences between our approach and the bookmaker are substantial. Based on our model, Denmark is 
the clear favorite for becoming IHF World Champion~2019.
These discrepancies could be mostly explained by the fact that 
the Lasso coefficient estimates from Table~\ref{tab:lasso_coefs}
include several other covariate effects beside the bookmaker's odds.

\subsection{Group rankings}

Finally, based on the 100,000 simulations, we also provide
for each team the probability to reach the main round. The results together 
with the corresponding probabilities are presented in Table~\ref{tab:group}.

Obviously, there are large differences with respect
to the groups' balances. While the model forecasts for example Spain and Croatia in Group~B, Denmark and Norway in Group~C and Hungary and Sweden in Group~D with probabilities clearly larger than $90\%$ to reach the second group stage, in Group~A France followed by Germany are the main favorites, but with lower probabilities
of $89.05\%$ and $80.16\%$, respectively. Hence, 
Group~A seems to be more volatile.


\begin{table}[h!]
\begin{center}
\small
 \caption{Probabilities for all teams to reach the main round at the IHF World Cup 2019 based on 100,000 simulation runs.}\label{tab:group}\vspace{0.4cm}
\begin{tabular}{|cr||cr||cr||cr|}
  \bottomrule
\multicolumn{2}{|c||}{\parbox[0pt][1.5em][c]{0cm}{} Group A}  &  \multicolumn{2}{|c||}{Group B}  &  \multicolumn{2}{|c||}{Group C}  &  \multicolumn{2}{|c|}{Group D} \\
\toprule  \bottomrule
  &&&&&&&  \\
    1. \cellcolor{lightgray} \includegraphics[width=0.4cm]{FRA.png}\enspace FRA  & \cellcolor{lightgray}$89.05\%$ &
    1. \cellcolor{lightgray}\includegraphics[width=0.4cm]{ESP.png}\enspace  ESP & \cellcolor{lightgray}$95.96\%$ &
    1. \cellcolor{lightgray}\includegraphics[width=0.4cm]{DEN.png}\enspace  DEN & \cellcolor{lightgray}$99.36\%$ &
    1. \cellcolor{lightgray}\includegraphics[width=0.4cm]{HUN.png}\enspace  HUN & \cellcolor{lightgray}$95.52\%$\\
&&&&&&&    \\[-3pt]

    2. \cellcolor{lightgray}\includegraphics[width=0.4cm]{GER.png}\enspace GER  & \cellcolor{lightgray}$80.16\%$ &
    2. \cellcolor{lightgray}\includegraphics[width=0.4cm]{CRO.png}\enspace  CRO & \cellcolor{lightgray}$91.8\%$ &
    2. \cellcolor{lightgray}\includegraphics[width=0.4cm]{NOR.png}\enspace  NOR & \cellcolor{lightgray}$93.49\%$ &
    2. \cellcolor{lightgray}\includegraphics[width=0.4cm]{SWE.png}\enspace  SWE & \cellcolor{lightgray}$93.81\%$ \\
&&&&&&&   \\[-3pt]
    3. \cellcolor{lightgray}\includegraphics[width=0.4cm]{RUS.png}\enspace RUS & \cellcolor{lightgray}$68.99\%$ &
    3. \cellcolor{lightgray}\includegraphics[width=0.4cm]{ICE.png}\enspace  ICE & \cellcolor{lightgray}$80.16\%$ &
    3. \cellcolor{lightgray}\includegraphics[width=0.4cm]{AUT.png}\enspace  AUT & \cellcolor{lightgray}$53.75\%$ &
    3. \cellcolor{lightgray}\includegraphics[width=0.4cm]{EGY.png}\enspace  EGY & \cellcolor{lightgray}$46.52\%$ \\
&&&&&&&   \\[-3pt]
    4. \cellcolor{lightgray}\includegraphics[width=0.4cm]{SRB.png}\enspace SRB & \cellcolor{lightgray}$52.83\%$ &
    4. \cellcolor{lightgray}\includegraphics[width=0.4cm]{MAC.png}\enspace  MAC & \cellcolor{lightgray}$21.38\%$ &
    4. \cellcolor{lightgray}\includegraphics[width=0.4cm]{TUN.png}\enspace  TUN & \cellcolor{lightgray}$50.66\%$ &
    4. \cellcolor{lightgray}\includegraphics[width=0.4cm]{ARG.png}\enspace  ARG & \cellcolor{lightgray}$28.45\%$ \\
&&&&&&&   \\[-3pt]
    5. \cellcolor{lightgray}\includegraphics[width=0.4cm]{BRA.png}\enspace BRA & \cellcolor{lightgray}$7.91\%$ &
    5. \cellcolor{lightgray}\includegraphics[width=0.4cm]{JPN.png}\enspace  JPN & \cellcolor{lightgray}$10.5\%$ &
    5. \cellcolor{lightgray}\includegraphics[width=0.4cm]{KSA.png}\enspace  KSA & \cellcolor{lightgray}$1.41\%$ &
    5. \cellcolor{lightgray}\includegraphics[width=0.4cm]{KAT.png}\enspace  KAT & \cellcolor{lightgray}$24.75\%$\\
&&&&&&&   \\[-3pt]
    6. \cellcolor{lightgray}\includegraphics[width=0.4cm]{KOR.png}\enspace KOR  & \cellcolor{lightgray}$1.06\%$ &
    6. \cellcolor{lightgray}\includegraphics[width=0.4cm]{BAH.png}\enspace  BAH & \cellcolor{lightgray}$0.2\%$ &
    6. \cellcolor{lightgray}\includegraphics[width=0.4cm]{CHI.png}\enspace  CHI & \cellcolor{lightgray}$1.34\%$ &
    6. \cellcolor{lightgray}\includegraphics[width=0.4cm]{ANG.png}\enspace  ANG & \cellcolor{lightgray}$10.95\%$ \\
&&&&&&&   \\[-3pt]
    \bottomrule
 \end{tabular}\vspace{0.4cm}
\end{center}
\end{table}



\section{Concluding remarks}\label{sec:conclusion}

In this work, we first compared four different regularized regression models for the scores of handball matches with regard to their predictive performances
based on all matches from the four previous IHF World Cups 2011 -- 2017, namely {\em (over- and underdispersed) Poisson regression models} and  {\em Gaussian response models}. 

We chose the Gaussian response model with constant 
and rather low variance (indicating a tendency of underdispersion) as our final model as the most promising candidate and fitted it to a training data set containing all matches of the four previous IHF World Cups 2011 -- 2017. Based on the corresponding estimates, we repeatedly simulated the IHF World Cup 2019 100,000 times. According to these simulations, the teams from Denmark ($37.2\%$) and France ($19.4\%$)
turned out to be the top favorites
for winning the title, with a clear advantage for Denmark.

\bibliographystyle{agsm}
\bibliography{literatur}

\end{document}